\documentclass[]{interact}

\usepackage[english]{babel}
\usepackage[utf8]{inputenc}
\usepackage[T1]{fontenc}
\usepackage{amssymb}
\usepackage{amsmath}
\usepackage{graphicx}
\usepackage{epstopdf}
\usepackage[labelfont=bf]{caption}
\usepackage{subcaption}
\usepackage[hidelinks=true]{hyperref}
\usepackage{cite}

\newcommand{\bigO}{\mathcal{O}}
\newcommand{\etal}{\textit{et al}.\ }
\newcommand{\mypar}[1]{\paragraph*{\textit{#1}.}}
\title{Implementation of Morse--Witten theory for a polydisperse wet 2D foam simulation}

\author{
  \name{F.F. Dunne$^*$\thanks{$^*$ Communicating author: dunneff@tcd.ie}, J. Winkelmann, D. Weaire, and S. Hutzler}
  \affil{School of Physics, Trinity College Dublin, University of Dublin, Ireland}
}

\date{}

\begin{document}
\maketitle


\begin{abstract}
  The Morse--Witten theory (D. Morse and T. Witten, EPL 22 (1993) 549--555)  provides a formulation for the inter-bubble forces and corresponding deformations in a liquid foam, accurate in the limit of high liquid fraction.
Here we show how the theory may be applied in practice, including allowing for polydispersity in the bubble sizes.
The resulting equilibrated 2D structures are consistent with direct calculations, within the limitations of the theory.
The path to developing a 3D model is outlined for future work.
\end{abstract}

\section{Introduction}
The theory of Morse and Witten  \cite{Morse93,WeaireHohlerReview} yields formulae relating forces, distortions, and energies of a bubble (or droplet in the analogous case of an emulsion), under the action of forces due to contacts with walls or other bubbles.
It proceeds from the case of a single bubble, pressed against a wall by buoyancy.
An extension to the case of multiple contacts (and hence a foam), also in static equilibrium, was indicated in the original paper \cite{Morse93}.
However, this has never been fully developed; for example, it does not account for polydispersity.
Hence, while there have been some limited trials of the theory  \cite{buzza1994uniaxial,HoehlerCohen-Addad2017} they have been restricted to monodisperse, or near monodisperse systems.
In the present paper we take some steps towards a full implementation of the theory of Morse and Witten allowing for an arbitrary degree of polydispersity.

The theory reduces a foam (or emulsion) to a set of representative points (the centres of mass of the bubbles) with central forces between them, depending on their separation but not as simple \emph{local} relations.
H\"ohler and Weaire  \cite{WeaireHohlerReview} have provided a review of the Morse--Witten theory, to which reference may be made for a more detailed understanding if necessary.

The present paper deals mainly with the case of a 2D foam, for which Weaire \etal  \cite{Weaire2017} have developed theory analogous to the original 3D case; this is the starting point for the present work.

The 2D foam, while not completely realised in practical systems (such as that of bubbles trapped between two plates) is a familiar test ground for the theory of foams \cite{CoxJaniaud08}.
Generally similar to 3D foam, in terms of its properties, it is simpler in many respects, and more readily simulated and visualised.
We anticipate analogous methods and results for the 3D foam, albeit with some important differences in detail, and a greater challenge to practical simulation.

\label{s:intro}
Relatively \emph{dry} (less than 10\% liquid) 2D foam has been successfully simulated in the past with the Plat software  \cite{Bolton91, Bolton92, BoltonWeaire90, PLAT, DunneEtal2017,Hutzler95}.
It is not based on an energy minimisation routine, but instead directly implements local equilibrium for a wet 2D foam.
It models the films and liquid-gas interfaces as circular arcs, constrained to meet smoothly at vertices.
This makes it quite an accurate model of 2D foam; however, the software suffers from a failure to converge for liquid fractions close to the wet limit.
Therefore, we seek a method for 2D foams that is successful in that limit.

\section{Morse--Witten theory in two dimensions}

\subsection{Basic Theory}
In the primitive version of the 2D theory, a 2D bubble is pressed against a fixed line by a buoyancy force \cite{Weaire2017}.
Just as in the 3D case, the distortion of the bubble shape from circular may be found in approximate analytic form, by solving a linearised Young--Laplace equation.
This solution can be used to build up a description of the foam of many bubbles, and the forces between them.
\begin{figure}
  \centering
  \includegraphics[
    width=0.6\textwidth
  ]{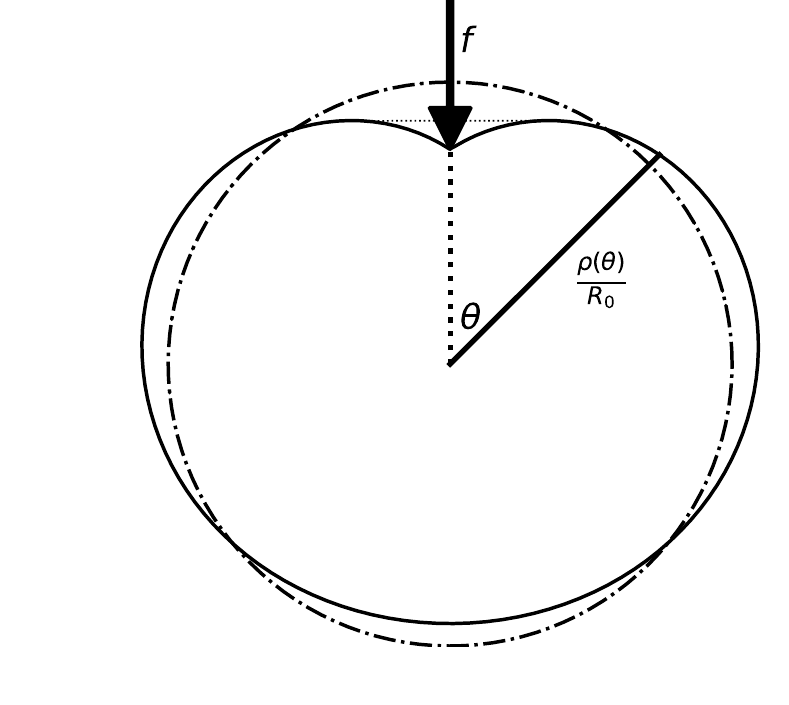}
  \caption{
    The profile $\rho(\theta)$ of a 2D bubble in terms of the polar angle $\theta$ under the action of a point force $f = F / \gamma = 1$, and an equal compensating body force, as calculated using Morse--Witten theory, Equations \eqref{e:profile} and \eqref{e:mwdeformation}.
    The undeformed circular bubble with radius $1$ is indicated by the dashed line.
    The part of the profile below the faint horizontal dashed line is disregarded.
  }
  \label{f:solo}
\end{figure}
It is expressed in terms of the radius $\rho(\theta)$, whose deviation from the unperturbed value $R_0$ is $\delta R(\theta)$,
\begin{equation}
  \rho(\theta)  = R_0 + \delta R(\theta),
  \label{e:profile}
\end{equation}
where $\theta$ is a polar angle relative to the point of contact.

The solution of the linearised Young--Laplace equation then results in  \cite{Weaire2017}
\begin{equation}
  \delta R(\theta)
  = \frac{R_0 F}{2 \gamma \pi} g(\theta)
  \text{,\quad with }
  g(\theta) =  (\pi - \theta) \sin (\theta)-\frac{\cos (\theta)}{2}-1 .
  \label{e:mwdeformation}
\end{equation}
Here $F$ is the magnitude of the total contact force, $\gamma$ is the line tension, and $g(\theta)$ encapsulated the deformation of a bubble in response to a force as in Weaire \etal \cite{Weaire2017}.
In the following we will often use the dimensionless force $f = F/\gamma$.

Equation~\eqref{e:mwdeformation} represents the deformation of the bubble in such a way that its centroid (or centre of mass), which represents its location, is kept fixed.
The profile $\rho(\theta)$ (Equation~\eqref{e:profile}) is shown in \figurename~\ref{f:solo}; it may be considered to represent a 2D bubble under the action of a \emph{point} force $F$ at $\theta = 0$, but can be used more generally.
Real bubbles would not support such a singular deformation.
Nevertheless, $g(\theta)$ can be used to predict the shape of a bubble subjected to realistic force distributions.

If this model is used to describe a contact with a straight line, analogous to a flat hydrophobic wall in 3D, then only part of this function is used, the profile being ``capped'' by a straight line \cite{WeaireHohlerReview}.
This is the only case considered (in 3D) by Morse and Witten: hence the previous restriction to monodisperse foams.
When describing polydisperse foams we require to deal with contacts with a curved boundary, appropriate to contacts between bubbles of different size (and pressure).

The reader unfamiliar with this subject may wonder why a body force (which we call buoyancy) has been introduced, while it has no place in the problem posed (a foam in the absence of gravity).
In fact, the solution for a bubble under the action of several forces \emph{in equilibrium} may be developed as a combination of the solution given here for the contacts of each bubble, with the effect of body forces cancelling out \cite{WeaireHohlerReview}.

\subsection{Contact between two bubbles of different sizes}
\label{s:two_bubbles}
\begin{figure}[t]
  \centering
  \includegraphics[
    width=0.95\textwidth
  ]{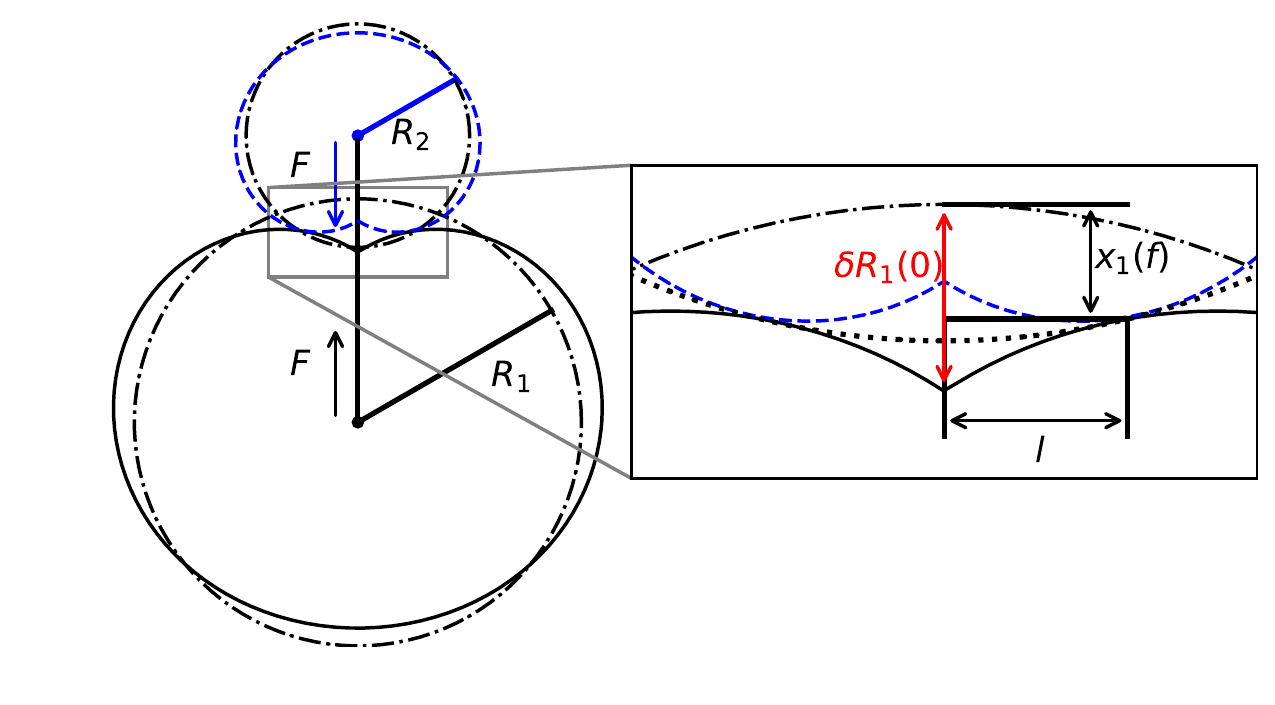}
  \caption{
    Two different sized 2D bubbles held in contact with each other by opposed body forces $F$, as calculated using Equations \eqref{e:profile} and \eqref{e:mwdeformation}.
    Their undeformed circular form with radii $R_1$ and $R_2$ is again illustrated by the dashed lines.
    Here we have used a large force for illustrative purposes; the theory is not accurate for deformations as large as this.
    Note the significance of the deformation $x_i$ (Equation~\eqref{e:polydisperse}), here illustrated for the bubble of radius $R_1$.
  }
    \label{f:2bubbles}
\end{figure}

\begin{figure}
  \centering
   \includegraphics[width=\textwidth]{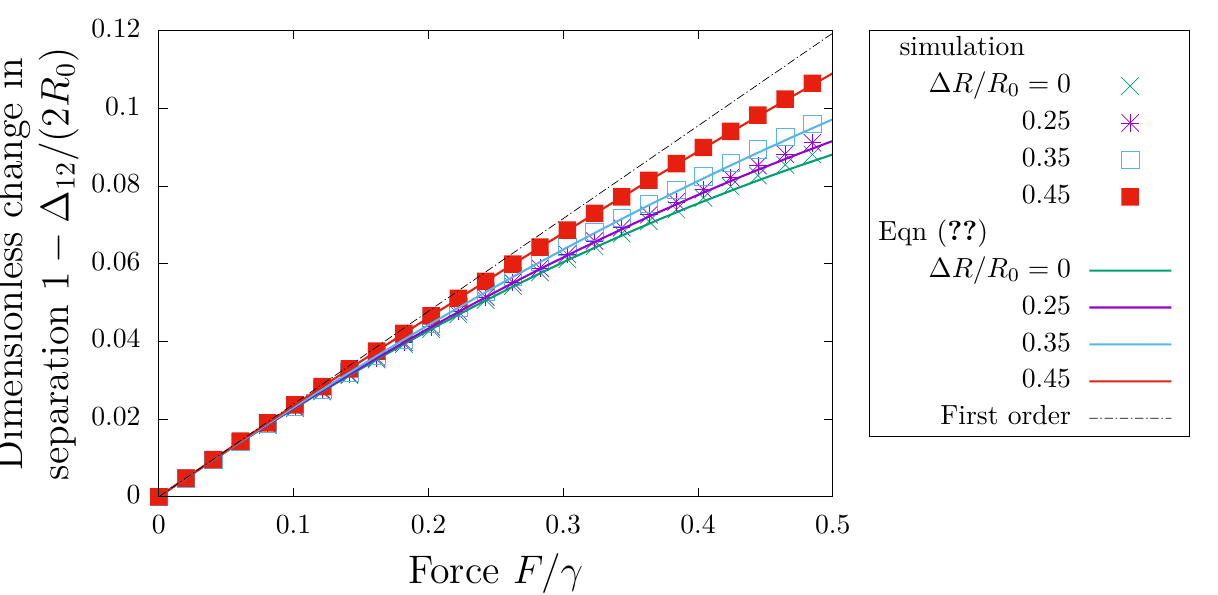}
    \caption{
      Dimensionless change in separation $1 - \Delta_{12}/(2R_0)$ versus force $f = F/\gamma$ between two 2D Morse--Witten bubble profiles (\figurename~\ref{f:2bubbles}) for varying size difference $\Delta R / R_0$.
      Symbols refer to numerical results from moving overlapping bubbles apart (see text), solid lines to Equation~\eqref{e:separation_change}.
      Up to a normalised force of $0.5$ the relative error of the theory is less than 2\%.
  }
  \label{f:overlap}
\end{figure}

Here we provide a generalisation of the Morse--Witten method to account for contacts between 2D bubbles of different sizes.
We require to find the relation between the contact force and the deformation of each bubble, represented by $x_i$, i.e. the distance along the centre--centre line from the undeformed bubble to the contact point (see the inset of \figurename~\ref{f:2bubbles}).
To lowest order, $x_i$ is the distance that the point at the cusp of the contact is displaced, which is
\begin{equation}
  - \delta R_1(0)
  =
  3 R_1 f/4 \pi
  \label{e:first_order_deformation}
\end{equation}
from Equation~\eqref{e:mwdeformation}.
This is indicated in red in the inset of \figurename~\ref{f:2bubbles}.
However, this overestimates the deformation at the contact (see \figurename~\ref{f:overlap}).

A simple derivation of the required relation between $F$ and $x_i$ follows.
As with many other aspects of the theory, this deals with lowest-order expressions only, and can be developed most expeditiously by using these from the outset (and verifying by a more cautious method if necessary).
Thus we can take for the force between two bubbles, to lowest order,
\begin{equation}
  F = 2l p_0
\end{equation}
where $2l$ is the width of the contact (\figurename~\ref{f:2bubbles}) and $p_0$ is the mean of the two (lowest order) bubble pressures $p_i = \frac{\gamma}{R_i}$.
Hence
\begin{equation}
  F
  \simeq
  \theta_i R_i \gamma \left( \frac{R_1 + R_2}{R_1 R_2} \right),
\end{equation}
where $2\theta_i$ is the opening angle of the contact, for $i=1,2$.
We can also express $x_i$ in terms of $\theta_i$, as
\begin{equation}
  x_i = R_i - \cos{(\theta_i)}\rho(\theta_i) \simeq -\delta R_i(\theta_i)
  \label{e:deformation}
\end{equation}

%
%
This improved expression for the deformation of bubble $i$ (= 1 or 2) is then expanded to $\bigO(f^2)$ to give
\begin{equation}
  x_i(f) =
    \frac{3 f R_i}{4 \pi }
    - \frac{f^2 R_i}{2(2 + R_1/R_2+ R_2/R_1)}
    ,
    \label{e:polydisperse}
\end{equation}
which is indicated in the inset of \figurename~\ref{f:2bubbles}.
The relative deformation $x_i/R_i$ is the same for each of the two bubbles.
The centre--centre distance $\Delta_{12}$ is then given by
\begin{equation}
  \Delta_{12} = (R_1 - x_1(f)) + (R_2 - x_2(f)).
\end{equation}
For two bubbles with radii $R_1 = R_0 + \Delta R$ and $R_2 = R_0 - \Delta R$, this results in the dimensionless change in separation as
\begin{equation}
  1 - \frac{\Delta_{12}}{2 R_0}
  =
  \frac{
    x_1(f) + x_2(f)
  }{
    2 R_0
  }
  =
  \frac{3 f}{4 \pi }
  -\frac{f^2}{8}
  \left[
    1
    -\left(
      \frac{\Delta R}{R_0}
    \right)^2
  \right]
  .
  \label{e:separation_change}
\end{equation}
Thus, terms of order $f^2$ or higher need to be considered in the expansion of Equation~\eqref{e:deformation} to account for polydispersity,
This is in contrast to the situation in 3D, see Section~\ref{s:3d}.
(Note that, in the monodisperse case, i.e. $\Delta R = 0$, the correction term in Equation~\eqref{e:separation_change} reduces to $f^2/8$, not to $f^2/4$, as erroneously stated in the appendix of Weaire \etal \cite{Weaire2017}.
This had no consequences for the results presented in that paper.)


In order to test the accuracy of Equation~\eqref{e:polydisperse} we proceed as follows.
For a given force, $f$, we draw two overlapping bubbles with facing contacts, using Equations~\eqref{e:profile} and \eqref{e:mwdeformation}.
The centres of these are moved apart until their area of overlap is zero, giving the separation for that force.

For the range of normalised force shown ($0 < f < 0.5$), Equation~\eqref{e:separation_change} produces a relative error $<2\%$ (the relative error when considering only its linear part is up to $25\%$, see \figurename~\ref{f:overlap}).

Equation~\eqref{e:polydisperse} may also be used to describe the case of a bubble $i$ in contact with \emph{multiple} bubbles (of different radii).
This results in a set of deformations at its contacts with its neighbours, $j$.
The deformation $x_{ij}$ of bubble $i$ due to its contact with bubble $j$, is determined by the sum over \emph{all} the contacts of bubble $i$,
\begin{equation}
  x_{ij} = - \frac{R_i}{2 \pi \gamma} \left(
    \sum_{k}^{N} F_{ik} g(\Delta \theta_{jk})
  \right)
  - \frac{R_i F_{ij}^2}{2 \gamma^2(2 + R_i/R_j + R_j/R_i)}.
  \label{e:deformationCalc}
\end{equation}
Here $\Delta \theta_{jk}$ is the angle the between the centre-centre lines of bubbles $i$ to $j$ and bubbles $i$ to $k$, where $k$ enumerates all the contacts of bubble $i$ (including $j$).
$F_{ij}$ is the force experienced by bubble $i$ at its contact with bubble $j$.

The equivalent expression for three dimensions is given in Section~\ref{s:3d}.

Note however that the linearised theory contains errors of order $f^2$ from the outset which we do not claim to eliminate.
Given that, the theory is surprisingly successful in improving the lowest order estimate.
The situation is similar to that which was encountered in the application of Morse--Witten theory to the pendant drop, although different in detail \cite{hutzler2018}.

\section{Formulation of the Morse--Witten model}
\label{s:conditions}

\subsection{Description of a foam}
We will proceed to apply Equation~\eqref{e:deformationCalc} to find an equilibrium structure of a polydisperse foam, in a numerical simulation.
We consider $N$ bubbles in equilibrium in a square box with periodic boundary conditions.
The bubbles are represented by their centroid positions ($\mathbf{c}_i$) and radii $R_i$.
A contact between bubbles $i$ and $j$ has an associated contact force of magnitude $F_{ij}$.

This nonlinear problem is naturally approached by iterative methods.
While its defining equations are simple, its implementation is challenging, because of the role of the contact network, which needs to be continually monitored and updated, as explained below.

\subsection{Defining Equations}
We seek an equilibrium configuration which satisfies the conditions \textit{A-D} below where the variables to be yielded by iteration are
\begin{itemize}
  \item the centre positions $\mathbf{c}_i$,
  \item the contact force magnitudes $F_{ij}$,
  \item and the contact deformations $x_{ij}$.
\end{itemize}

\mypar{A) Force-deformation relation}
Forces and deformations must be consistent, that is, satisfy Equation~\eqref{e:deformationCalc}.

\mypar{B) Deformation-displacement relations}
For each contact, the separation of centres of mass, located at positions $\mathbf{c}_i$ and $\mathbf{c}_j$ must be consistent with the deformations $x_{ij}$ and $x_{ji}$, according to
\begin{equation}
  R_i - x_{ij} + R_j - x_{ji}
  =
  |\mathbf{c}_i - \mathbf{c}_j|
  .
  \label{e:deformationConsistency}
\end{equation}
%
\mypar{C) Action-reaction}
\begin{equation}
  F_{ij} = F_{ji}
  \label{e:forceBalance}
\end{equation}
\mypar{D) Equilibrium of forces}
The vector net forces on each bubble $i$ must satisfy
\begin{equation}
  \sum_j^N F_{ij} \frac{\mathbf{c}_j - \mathbf{c}_i}{|\mathbf{c}_j - \mathbf{c}_i|} = 0
  .
  \label{e:equilibrium}
\end{equation}

\subsection{The contact network}
As the system approaches equilibrium  the shapes and positions of bubbles change.
The contact network is not finally determined until equilibrium is reached, consistent with the above conditions.
It requires to be updated as the approach to equilibrium proceeds.
Buzza and Cates  \cite{buzza1994uniaxial} applied the Morse--Witten theory to the case of an emulsion where the drops are arranged on a simple cubic lattice, for which this difficulty does not arise.
H\"ohler and Cohen-Addad  \cite{HoehlerCohen-Addad2017}, while including a slight polydispersity, also used crystalline systems in which contact changes were excluded.
For the disordered foams discussed here a new methodology is thus needed to deal with bubble rearrangements (topological changes).

\section{Implementation of the Morse--Witten model}
\subsection{Iterative scheme}
We have developed a practical iterative scheme that can produce an equilibrium structure satisfying the conditions of Section~\ref{s:conditions}.
Separate steps of iteration are designed to bring the configuration closer to satisfaction of the conditions.
To start a set of bubble centres and radii is required.
These can be obtained from other software, such as Plat \cite{PLAT}, Bubble model \cite{Durian95}, or Surface Evolver \cite{Brakke92}.

\mypar{A) Force-deformation relation}
Given a configuration and deformations of each bubble contact, the corresponding forces are found by solving Equation~\eqref{e:deformationCalc} for $F_{ik}$, for each bubble in turn.
This is a nonlinear equation, hence we solve it iteratively.
This difficulty is also to be found in the work of H\"ohler and Cohen-Addad \cite{HoehlerCohen-Addad2017}, and we adopt the same method as was used by them.
That is, in each iteration $n$, the forces from the previous iteration are inserted in the quadratic term, leaving a linear equation to be solved.
Additionally, we apply some damping to this procedure, implemented as
\begin{equation}
  F^{(n+1)} = aF^{(n+1)} + (1-a)F^{(n)}
\end{equation}
where we have found $a = 0.9$ to be a good choice.
This helps to prevent oscillations in the forces, without slowing convergence too much.

\mypar{B) Deformation-displacement relations}
The deformations are updated by
\begin{equation}
  x^{(n+1)}_{ij} = x^{(n)}_{ij} + \frac{R_j}{R_i + R_j}\Big[(R_i - x^{(n)}_{ij} + R_j - x^{(n)}_{ji}) - |\mathbf{c}_j - \mathbf{c}_i|\Big]
  \label{e:deformationUpdate}
\end{equation}
in order to satisfy Equation~\eqref{e:deformationConsistency}.

\mypar{C) Action-reaction}
$F_{ij}$ and $F_{ji}$ are replaced by their average.

\mypar{D) Equilibrium of forces}
Each bubble located at position $\mathbf{c}_i$ is moved in the direction of the net force acting on it, according to
\begin{equation}
  \mathbf{c}_i^{(n+1)} = \mathbf{c}_i^{(n)} + b\sum_j^N F_{ij}^{(n+1)} \frac{\mathbf{c}_i - \mathbf{c}_j}{|\mathbf{c}_i - \mathbf{c}_j|},
  \label{e:moveCenters}
\end{equation}
where $b = 0.1 R_0/\gamma$.
As convergence speed is directly proportional to $b$, we have selected as large a $b$ as possible for which the algorithm still converges.

\begin{figure}[p]
  \centering
  \includegraphics{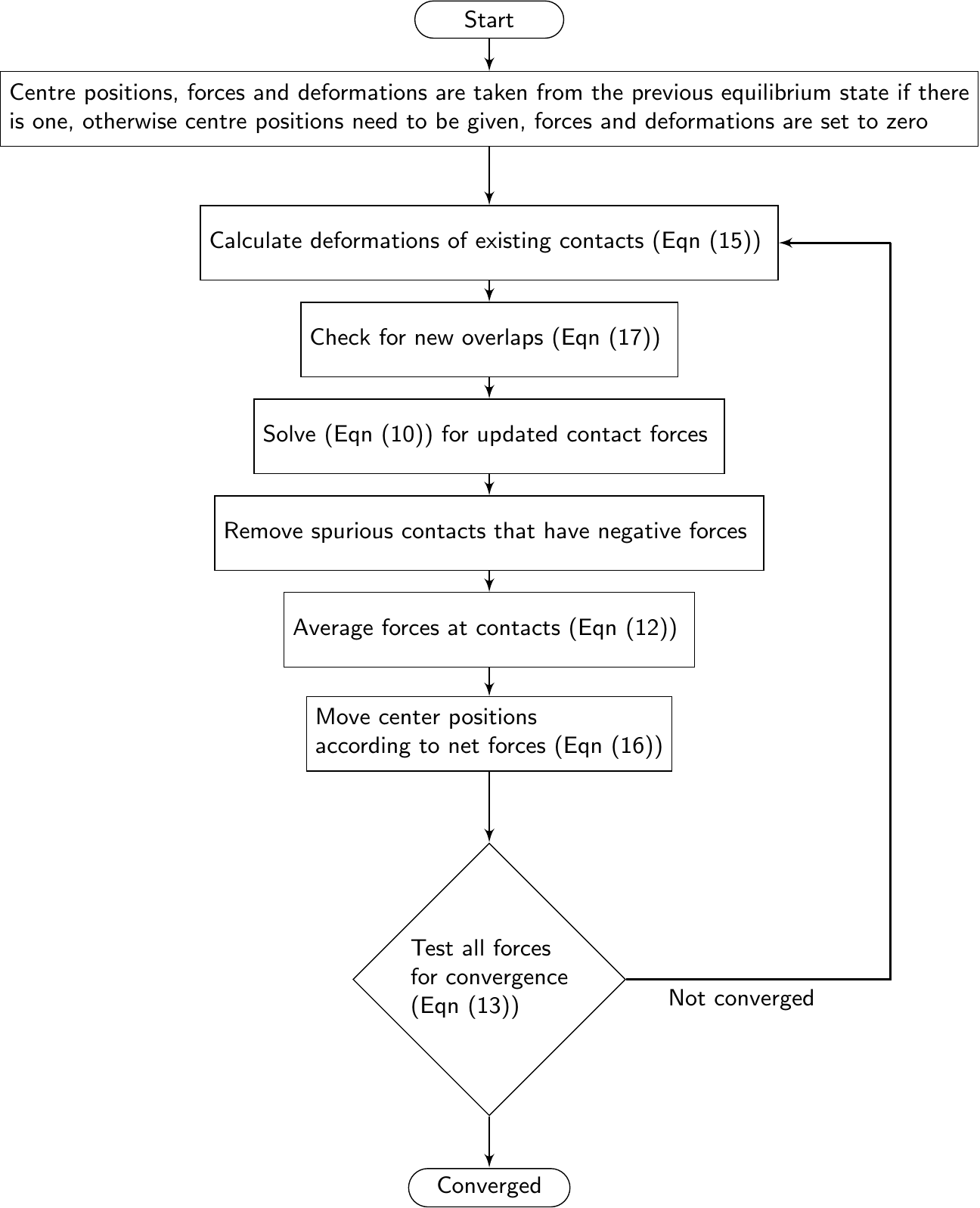}
  \caption{
    Iteration scheme for the computation of a  2D Morse--Witten foam.
    While the test forces are not converged, deformations, overlaps, and contact forces are calculated and the centroid positions moved accordingly.
    For a given collection of bubbles in a given confinement this procedure can be use to find an equilibrium configuration.
  }
  \label{f:flowchart}
\end{figure}

The flowchart of the iteration is shown in \figurename~\ref{f:flowchart}.
Note that it contains additional steps in which the contact network is, if necessary, altered.

\subsection{Updating the contact network}

\subsubsection*{Negative forces}
A negative force indicates a spurious contact, i.e. a contact which has arisen from an overlap while the system is out of equilibrium, and this is removed from the contact network.
In practice, at most one negative force is eliminated for each bubble in a given iteration to provide stability of the algorithm.
This is performed after updating the forces.

\subsubsection*{Overlapping bubbles}
After moving the bubble positions, nominally non-contacting bubbles may overlap with each other.
To detect this we calculate
\begin{equation}
  x_{ij} = \frac{R_j}{R_i + R_j}\Big[\rho(\theta_{ij}) + \rho(\theta_{ji}) - |\mathbf{c}_j - \mathbf{c}_i|\Big]
  \label{e:contactDetect}
\end{equation}
For $x_{ij} > 0$ bubbles $i$ and $j$ overlap.
This requires an update of the contact network, which is performed before updating the forces.

\subsection{Convergence}
The algorithm is terminated when the foam being simulated is close to equilibrium, satisfying all of the above requirements.
This is determined numerically by calculating the net force on each bubble in the foam using the left hand side of Equation~\eqref{e:equilibrium}.
We deem this equation to be satisfied for all bubbles if the largest net force encountered is less than $\gamma \times 10^{-4}$.

In this case the centroid positions given by the recurrence relation, Equation~\eqref{e:moveCenters}, will have converged, leading also to a convergence of the deformations, Equation~\eqref{e:deformationUpdate}.
Thus, solving the deformation-force relationship Equation~\eqref{e:deformationCalc} repeatedly will produce the same set of contact forces each time and all the defining equations will be satisfied.

\section{Tests and Typical Results}

We have run tests of the above scheme for systems of up to 200 bubbles (the run time of the program scales quadratically with the number of bubbles), in a square box with periodic boundary conditions.
The computations converged satisfactorily for liquid fraction exceeding around $\phi = 0.12$, a liquid fraction where at least 80\% of the Plat simulations fail \cite{DunneEtal2017}.
We have not identified the reason for non-convergence beyond that point, but it is hardly surprising in a nonlinear problem of this kind, and may be rectified in due course.
In order to validate the method, we have compared it with simulations  using the Plat software, as introduced in Section~\ref{s:intro}.

\begin{figure}[p]
  \centering
  \includegraphics[width=\textwidth]{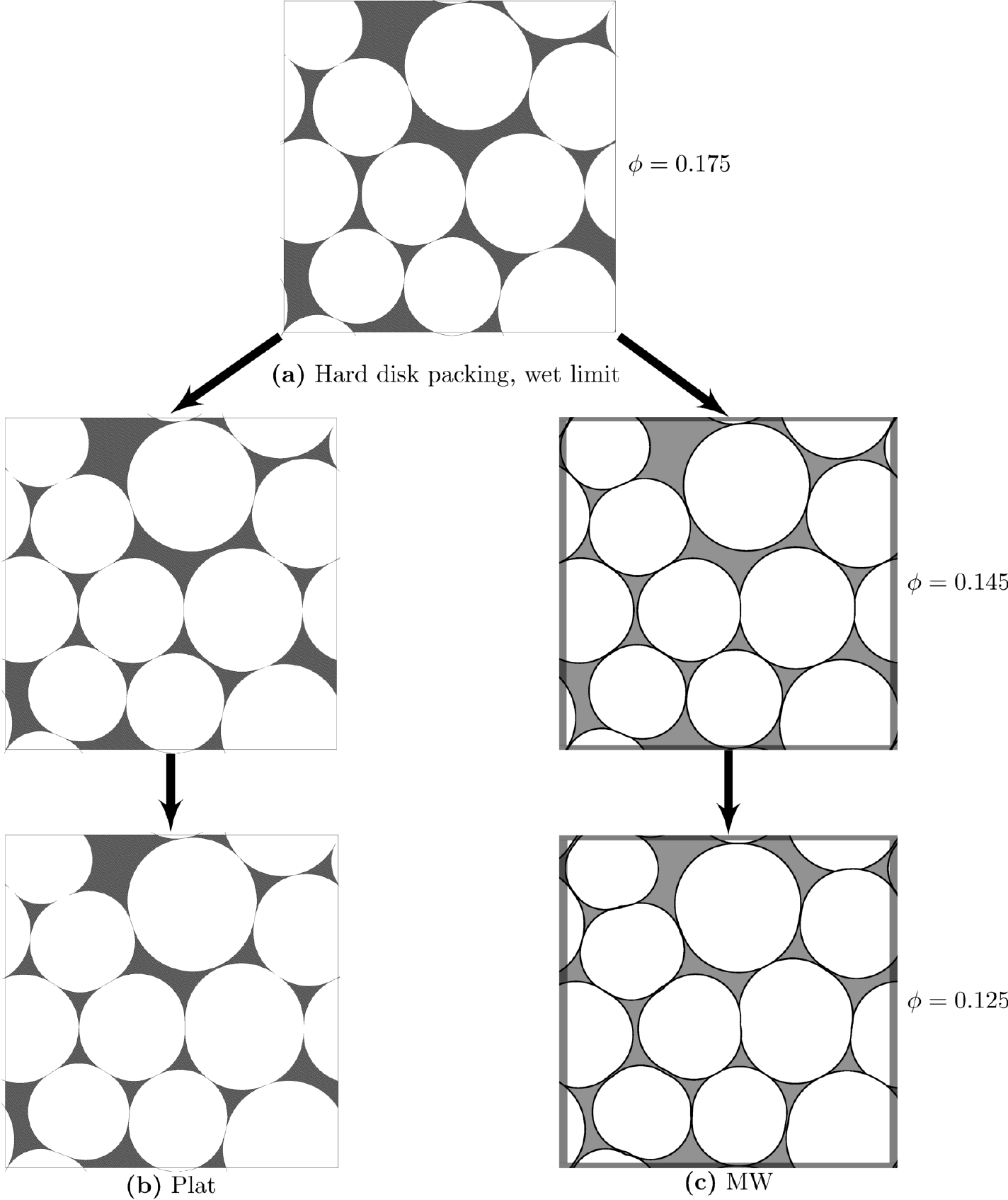}
  \caption{
    Comparison of polydisperse 2D foam as computed using the Plat simulation software  \cite{PLAT} and the Morse--Witten formulation.
    Each structure is derived from the same hard disk packing (a), by gradually decreasing the liquid fraction in steps of $\Delta\phi = 0.001$.
    The two simulation methods produce almost the same sequence of contact changes.
  }
  \label{f:compare}
\end{figure}

To begin, a system of ten bubbles with a polydispersity of $\sqrt{\langle R^2\rangle/\langle R\rangle^2 -1} \simeq 0.12$ was generated using the Plat software and the liquid fraction  $\phi$ increased until a hard disk packing was achieved (\figurename~\ref{f:compare}(a), top).
A hard disk packing corresponds to a foam in the wet limit (at $\phi = \phi_c$) where all of the degrees of freedom are exactly taken up by the contacts between bubbles, and there are no additional constraints.
In this case the average number of contacts is $Z_c = 4(1-1/N) = 3.6$ \cite{WinkelmannEtal2017jamming}.
Ten bubbles constitutes a small enough system that, despite the general failure of Plat to converge in the wet limit, the cost of repeating simulations until it is found to be successful is sufficiently small so as to make it feasible.
The centre positions of the bubbles were extracted and used to create a Morse--Witten simulation of the same system.
The liquid fraction of both simulations was then decremented in parallel, down to a liquid fraction of approximately 0.12.
The Morse--Witten simulation produced almost the same contact changes as the Plat simulation, although at values of $\phi$ shifted by roughly $\Delta \phi \simeq 0.01$ higher.
In looking at this comparison, it should be borne in mind that the Morse--Witten formalism is inherently approximate.

We next consider the excess energy of a Morse--Witten foam, defined (in dimensionless form) by
\begin{equation}
  \varepsilon
  =
  \frac{1}{ 4 \pi R_0 \gamma }
  \sum_{i=0}^N \sum_j x_{ij} F_{ij},
  \label{e:energy}
\end{equation}
where $j$ enumerates the contacts of bubble $i$.

\begin{figure}
  \centering
  (a)
  \includegraphics[width=0.8\textwidth]{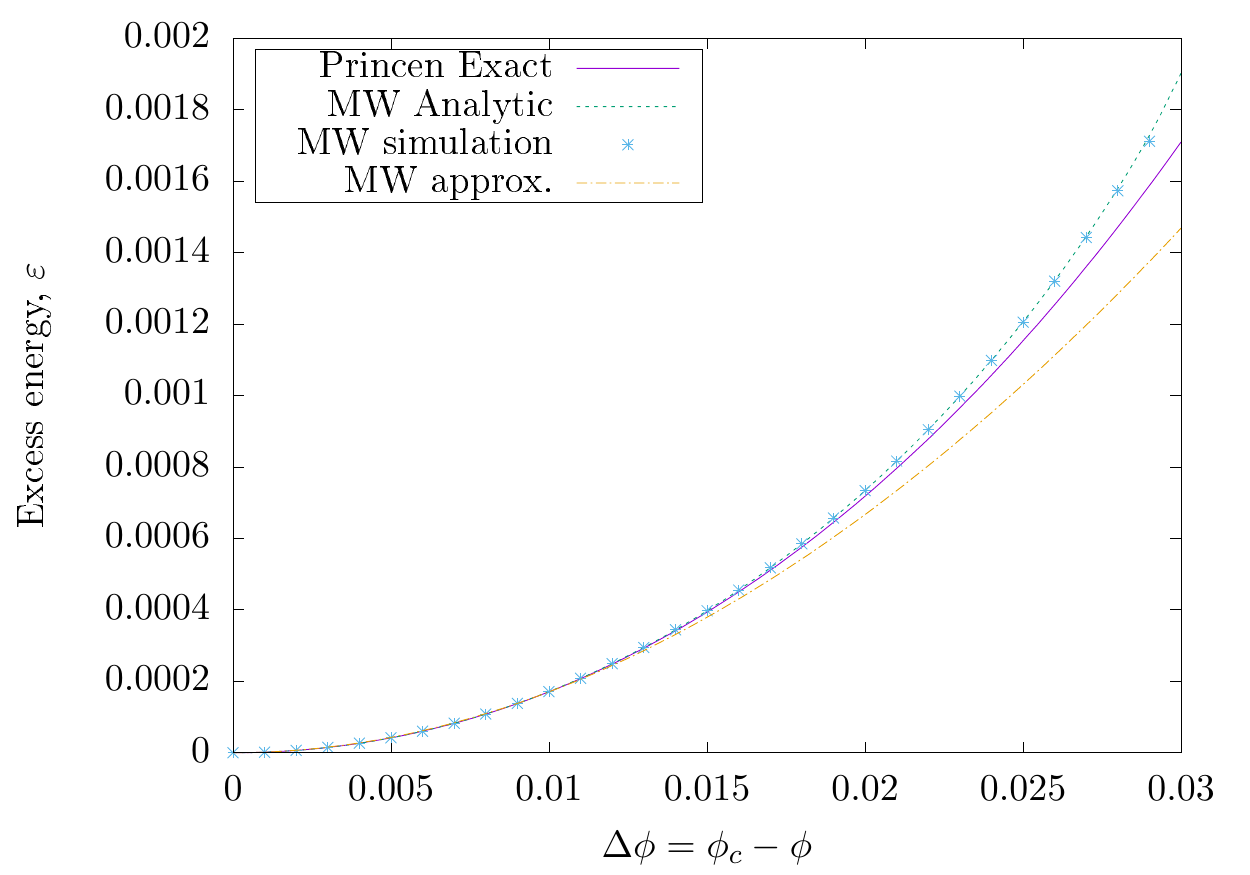}\\
  (b)
  \includegraphics[width=0.8\textwidth]{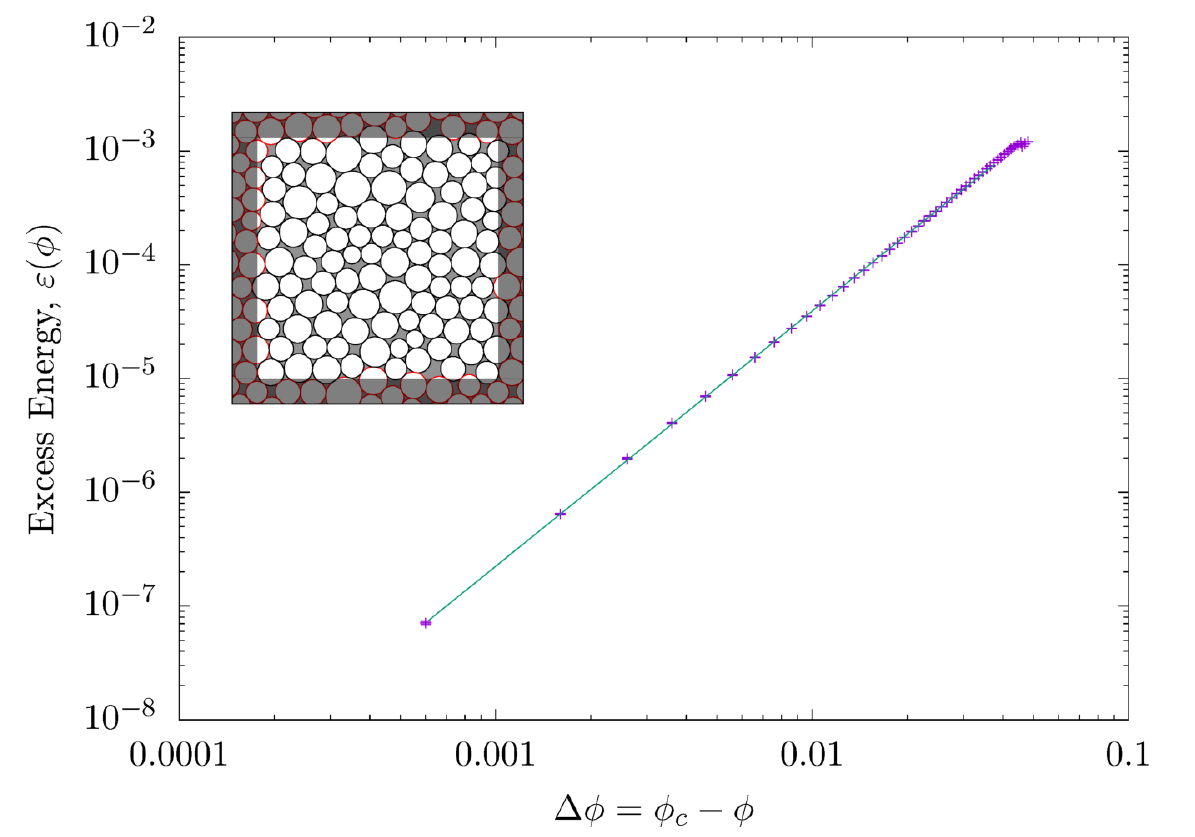}
  \caption{
    Variation of normalised excess energy $\epsilon$ (Equation~\eqref{e:energy}) as a function of excess liquid fraction $\Delta \phi = \phi_c - \phi)$.
    (a) In the case of an ordered monodisperse foam the Morse--Witten theory reproduces the exact result first derived by Princen \cite{contactAnglePaper,Princen1979} (data points: simulation, dashed line: analytic).
    Also shown is a simple analytic approximation obtained from Morse--Witten theory (Equation~\eqref{e:energy_calc}) (dot-dashed line).
    (b) For disordered foams, our simulations of 1000 systems of 100 bubbles each show that the excess energy is proportional to $\Delta \phi^{2.2}$.
    An example of one of the simulated foams is shown in the inset.
  }
  \label{f:energy}
\end{figure}

For ordered monodisperse foam, Princen calculated  $\varepsilon(\phi)$ exactly \cite{Princen1979,contactAnglePaper}.
This presents a good test for the Morse--Witten model, which can be solved exactly in this case.
\figurename~\ref{f:energy}(a) shows excellent agreement between Princen's exact result and the analytic solution of the Morse--Witten model in the wet limit ($\Delta \phi < 0.02$).
Our numerical simulation results match the analytic solution of the Morse--Witten model.

Also shown is a simple approximate solution of the Morse--Witten model, which can be obtained as follows.
The energy per contact is given by elementary methods as $0.5 F \delta R / (\gamma R_0)$ and using the relation
\begin{equation}
  \frac{F}{\gamma} = \frac{6 Z}{\pi} \frac{\delta R}{R_0}
  \label{e:deltar_f}
\end{equation}
from Weaire \etal \cite{Weaire2017}, along with the affine compression relation $\delta R / R_0 = \Delta \phi / 2 (1 - \phi)$, we obtain
\begin{equation}
  \varepsilon(\Delta \phi)
  =
  \left(
  \frac{3}{\sqrt{2 \pi}} \frac{\Delta \phi}{\Delta \phi + \phi_h}
  \right)^2
  ,
  \label{e:energy_calc}
\end{equation}
where $\phi_h = \pi/ 2\sqrt{3}$ is the critical packing fraction for a hexagonal disk arrangement.
This relation (shown in \figurename~\ref{f:energy}(a)) is in excellent agreement with the result of Princen for $\Delta \phi < 0.015$.

In order to study the variation of excess energy $\varepsilon$ as a function of liquid fraction of polydisperse foams, 1000 foams of 100 bubbles each were prepared with an average polydispersity of $0.21\pm0.02$.
These simulations were run for a range of liquid fraction from $0.18$ to $0.12$ in steps of $0.001$.
They were started deliberately higher than the expected value of $\phi_c \simeq 0.16$ so that the transition from unjammed collection of disks to jammed foams will not be missed.
The critical value $\phi_c$, marks the onset of the excess energy.

Our simulations show that, similar to results from Plat \cite{DunneEtal2017}, close to $\phi_c$, the energy varies roughly quadratically with the distance $\Delta \phi = \phi_c - \phi$ from $\phi_c$.
Therefore, the values for $\phi_c$ were calculated individually for each 100 bubble system by fitting a straight line to the lowest eight points of the square root of the energy curve that were above $10^{-4}$.
The average value obtained from this procedure is $0.843\pm0.003$, consistent with previously published values for $\phi_c$ \cite{BoltonWeaire90,WinkelmannEtal2017jamming,bideau1984compacity,MajmudarEtal2007,Durian95,Sun2004}.
The energy curves for these simulations were shifted by their respective $\phi_c$ values, and then averaged with a bin width of $0.001$ in $\Delta \phi$ to smooth the data.
\figurename~\ref{f:energy}(b) shows that, based on our 1000 simulations, $\varepsilon(\Delta \phi) \propto \Delta \phi^{2.2}$.

\begin{figure}[h]
  \centering
  \includegraphics[width=0.8\textwidth]{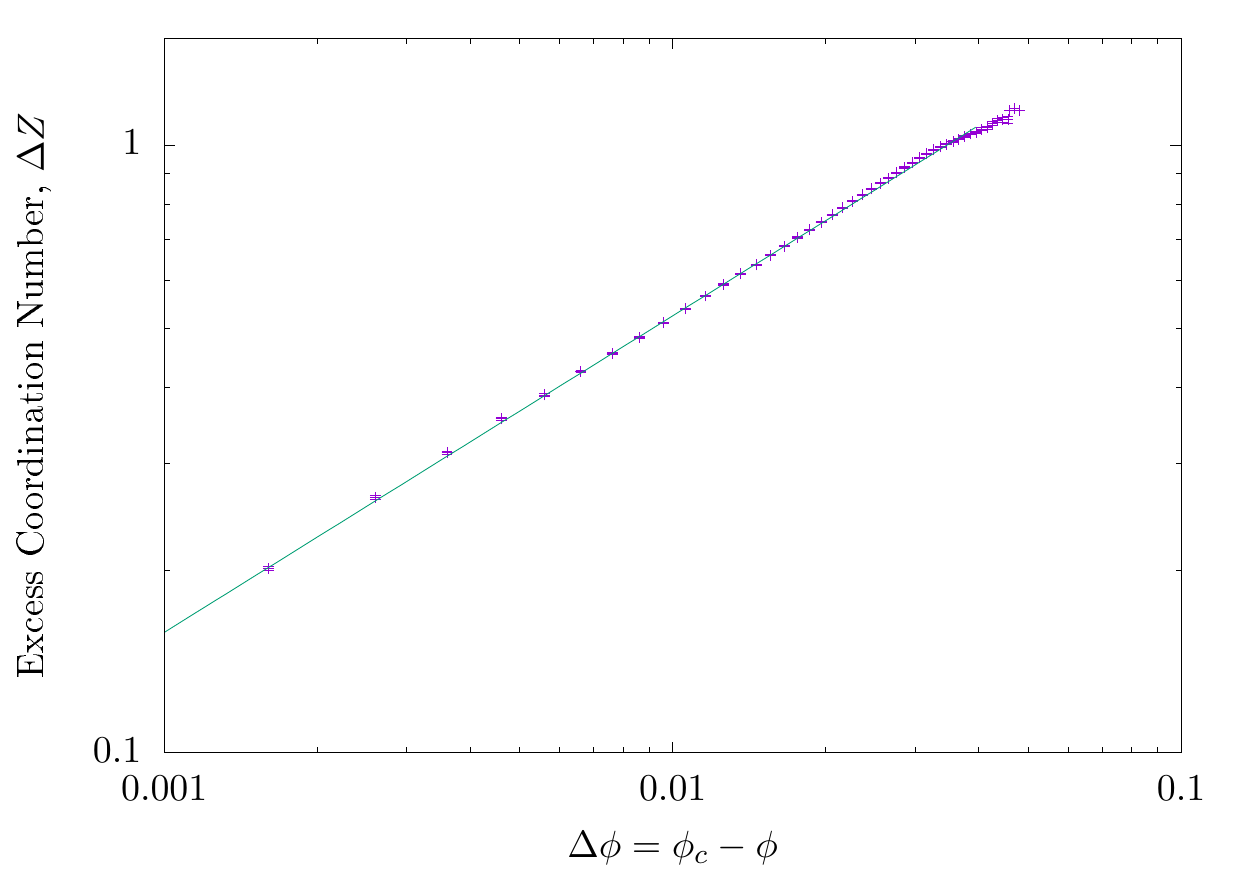}
  \caption{
    In the case of disordered foams, our simulations show an increase in the excess coordination number with excess liquid fraction of the form $\Delta Z \propto \Delta\phi^{0.52}$, consistent with previous simulations using the bubble model.
  }
  \label{f:z}
\end{figure}

A further quantity of interest in the context of random packings is the variation of the average coordination number, $Z$, with liquid fraction.
A log-log plot of our data (\figurename~\ref{f:z}) reveals a scaling of $Z - Z_c = \Delta Z \propto \Delta \phi^{0.52}$, consistent with results for packings using the soft disk model \cite{o2003jamming}.
Such a scaling was recently disputed based on extensive computer simulations with Plat which resulted in $\Delta Z \propto \Delta \phi$, and it was argued that this was due to the deformability of soft bubbles \cite{WinkelmannEtal2017jamming}.
The results presented here appear to put some doubts on this argument.
Further simulations with Plat and the Surface Evolver software \cite{Brakke92} (currently restricted to finite contact angles in two dimensions \cite{contactAnglePaper}) would be required to determine whether the reported linear scaling with $\Delta \phi$ might be due to some inherent bubble-bubble attraction that arises from the algorithms.

In the study of granular matter it is common to compute the contact force network \cite{Behringer1996,Snoeijer2004}.
Granular packings are characterised by a very slow decay of the distribution of forces greater than the mean.
Whether this is exponential or faster than exponential depends on the details of the simulations/experiments, such as dimensionality, solid friction, and partial size distribution \cite{vanEerd2007,Radjai1996,Liu1995}.

\begin{figure}
  \centering
    \includegraphics[width=\textwidth]{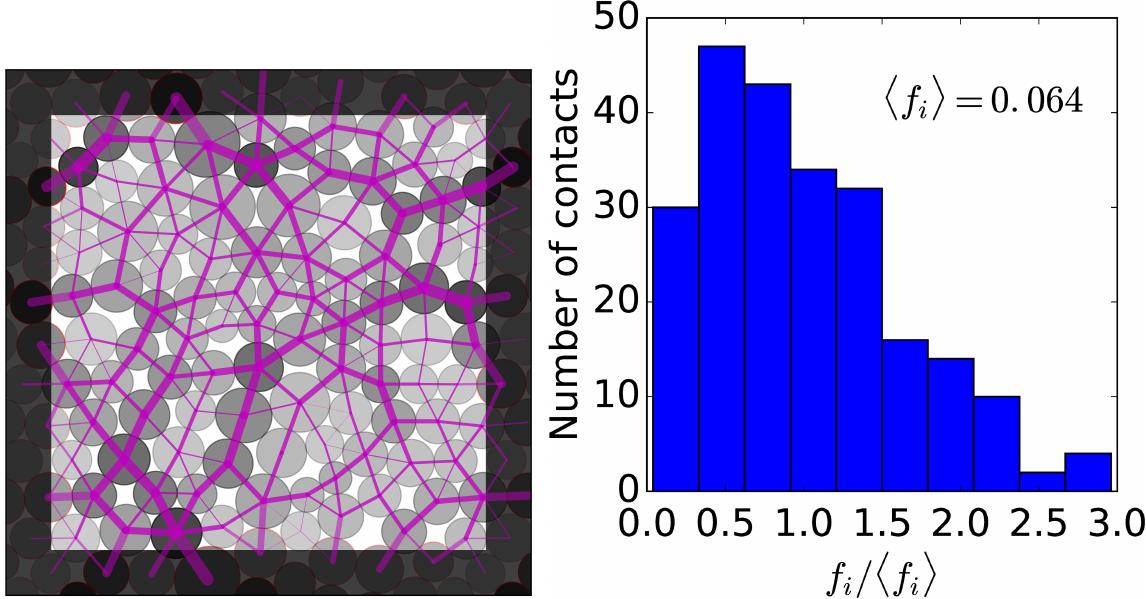}
  \caption{
    (Left) Wet foam ($\phi = 0.13$) with 100 bubbles showing the contact force network.
    The thickness of the lines is proportional to the force magnitude and the grey scale is proportional to the individual excess energy of a bubble.
    (Right) Normalised distribution of the forces. This is in qualitative agreement with that found by H\"ohler and Cohen-Addad \cite{HoehlerCohen-Addad2017}.
  }
  \label{f:large}
\end{figure}

In \figurename~\ref{f:large}(a) we show the contact force network for an equilibrated Morse--Witten foam of 100 bubbles at a liquid fraction of $\phi = 0.13$.
The width of each line in the contact network is proportional to the force magnitude.
In addition, the bubbles are shaded according to their individual excess energies.
Also shown in \figurename~\ref{f:large} is a preliminary normalised distribution of contact forces.
This is broadly similar to that found by H\"ohler and Cohen-Addad \cite{HoehlerCohen-Addad2017}, however, further simulations are required to analyse its shape.

\section{Extension to a 3D foam}
\label{s:3d}
The methodology developed above for the simulation of a 2D foam based on the Morse--Witten model lends itself to application also for 3D.
As in 2D the foam will be represented by the centroid of all bubbles and a network of contacts.
In 3D, the profile is expressed analogously to Equation~\eqref{e:profile} and Equation~\eqref{e:mwdeformation} becomes
\begin{equation}
  \delta R(\theta)
  =
  \frac{
    -F
  }{
    \gamma R_0
  }
  G(\theta)
\end{equation}
where
\begin{equation}
  G(\theta)
  =
  -\frac{1}{4 \pi }
  \left\{
    \frac{1}{2}
    +\frac{4}{3} \cos{\theta}
    +\cos{\theta} \ln{[ \sin^2(\theta/2) ]}
  \right\}
  ,
  \label{e:3dsolution}
\end{equation}
from \cite{Morse93}.
The expression for the deformation of bubble $i$, equivalent to Equation~\eqref{e:deformationCalc} and derivable in the same way, is given by
\begin{equation}
  x_{1}(F)
  =
  \frac{F} {4 \pi R_1 \gamma  } \left[
      \frac{11}{6}
      -\frac{2 R_2}{R_1 + R_2}
      +\ln \left(
        \frac{F  R_2}
        {
          4 \pi  R_1 \gamma  (R_1 + R_2)
        }
      \right)
    \right]
  ,
  \label{e:3d_defomation}
\end{equation}
to lowest order in $F$.
The relative change in separation (equivalent to Equation~\eqref{e:separation_change}) between two bubbles where $R_1 = R_0 + \Delta R$ and $R_2 = R_0 - \Delta R$ is
\begin{equation}
  1
  -\frac{\Delta_{12}}{2 R_0}
  =
  \frac{
    F
  }{
    4 \pi  \gamma  R_0
  }
  \left(
    \frac{5}{6}
    + \ln \left(
      \frac{F}{8 \pi  \gamma R_0}
    \right)
  \right)
  .
  \label{e:3d_separation}
\end{equation}
Again, symmetry tells us not to expect any terms of odd orders of $\Delta R$ in the separation.
Equation~\eqref{e:3d_defomation} would need to be expanded to order $F^2$ to give terms of $\Delta R^2$.
Taking for example $\Delta R = 0.1 R_0$, $F = 0.5 \gamma R_0$, the relative error that would result from using a formula for monodisperse foam would be of order $10^{-4}$.
This explains the success by H\"ohler and Cohen-Addad  \cite{HoehlerCohen-Addad2017} in using an expression derived for the monodisperse case in treating a slightly polydisperse case.

In order to model a 3D foam, an equivalent to Equation~\eqref{e:deformationCalc} is required.
This is obtained by adding a non-local term to Equation~\eqref{e:3d_defomation} (see \cite{HoehlerCohen-Addad2017}) giving
\begin{equation}
  x_{ij}(F)
  =
  \frac{F} {4 \pi R_i \gamma  } \left[
    \frac{11}{6}
    -\frac{2 R_j}{R_i + R_j}
    +\ln \left(
      \frac{F  R_j}
      {
        4 \pi  R_i \gamma  (R_i + R_j)
      }
    \right)
  \right]
  + \sum_{k \neq j} G(\Delta \theta_{jk}) \frac{F_{ik}}{R_i \gamma}
  .
\end{equation}

Similar to the procedure of Section~\ref{s:two_bubbles}, we determined the separation of two 3D Morse--Witten bubbles at their point of contact, for a given force $F$.
We find that Equation~\eqref{e:3d_separation} is reasonably accurate up to $F/(R\gamma) \sim 0.5$ (corresponding to the dry limit) for low polydispersity, and $F/(R\gamma) \sim 0.05$ (corresponding to $\phi \sim 0.24$) for high polydispersity.
The appearance of a non-linear $F\ln(F)$ term makes the 3D case somewhat different from the 2D one presented here: nevertheless we hope that further improvement of the 2D methods will assist in the greater computational task of implementation in 3D.

\section{Conclusion}
We have shown how polydispersity can be accommodated in the Morse--Witten theory, in such a way as to give satisfactory results for a typical disordered polydisperse foam that is close to the wet limit.
The extension of the theory to 3D is quite natural, although the implementation becomes conceptually more difficult to visualise and check, and there is an obvious increase in computational demands.
The transparency of the theory and its direct relation to a force network (\figurename~\ref{f:large}) is attractive.
However, it should be noted that it has proven a computational challenge that was hardly anticipated, and is worthy of further attention.

In the polydisperse foam the bubble-bubble interfaces have pronounced curvature: this is accounted for in the present formulation, being related to differences in bubble sizes.
One might well ask what is the case in a monodisperse \emph{disordered} foam?
(Despite some doubts in the past, this can indeed exist, even in 2D).
Since the bubbles are not equivalent, surely their pressures are slightly different, hence the interfaces are curved?
This is correct in principle, but the effect is surely very small, and of higher order in the forces than what is considered here.

\section{Acknowledgements}
DW wishes to acknowledge discussions with Reinhard H\"ohler and Tom Witten.

Research supported in part by a research grant from Science Foundation Ireland (SFI) under grant number 13/IA/1926 and from an Irish Research Council Postgraduate Scholarship (project ID  GOIPG/2015/1998).
We  also  acknowledge the COST action MP1305 ‘Flowing matter’ and the European Space Agency ESA MAP Metalfoam (AO-99-075) and Soft Matter Dynamics (contract: 4000115113).

\clearpage

\end{document}